\documentclass[runningheads]{llncs}
\usepackage{graphicx}
\usepackage{wrapfig}
\usepackage[hidelinks]{hyperref}

\usepackage{paralist}

\usepackage[disable]{todonotes}

\usepackage{booktabs}
\usepackage{tabularx}
\usepackage{soul} 

\usepackage{cite} 
\usepackage{multirow} 
\usepackage{makecell} 
\usepackage[binary-units=true]{siunitx}

\newcommand{\colfootnote}[1]{\unskip\kern-2pt\footnote{#1}}

\usepackage{xspace}
\newcommand{\eg}{\emph{e.g.}\xspace}
\newcommand{\ie}{\emph{i.e.}\xspace}
\newcommand{\etal}{\emph{et al.}\xspace}

\begin{document}

\title{Tell-Tale Tail Latencies: \break Pitfalls and Perils in Database Benchmarking}
%
%
\author{Michael Fruth\inst{1}\orcidID{0000-0003-2933-5093} \and
Stefanie Scherzinger\inst{1}\and \\
Wolfgang Mauerer\inst{2,3}\orcidID{0000-0002-9765-8313} \and
Ralf Ramsauer\inst{2}}
\authorrunning{M.~Fruth et al.}
%
\institute{University of Passau, 94032 Passau, Germany
\email{\{michael.fruth,stefanie.scherzinger\}@uni-passau.de}\\\and
Technical University of Applied Sciences Regensburg, 93058 Regensburg, Germany
\email{\{wolfgang.mauerer,ralf.ramsauer\}@othr.de}\\\and
Siemens AG, Corporate Research, Otto-Hahn-Ring 6, 81739 Munich, Germany}
\maketitle 
\begin{abstract}
The performance of database systems is usually characterised by their
average-case (\ie, throughput) behaviour in standardised or de-facto standard benchmarks 
like TPC-\emph{X} or YCSB. While tails of the latency (\ie, response time) distribution receive
considerably less attention, they have been identified as a threat to the overall system
performance: In large-scale systems, even a  
fraction of requests delayed can build up into delays perceivable by end users. 
To eradicate large tail latencies from database systems, the ability to
faithfully record them, and likewise pinpoint them to the root causes, 
is imminently required.  In this paper, we address the challenge of
measuring tail latencies using standard benchmarks, and identify
subtle perils and pitfalls.
In particular, we demonstrate how Java-based
benchmarking approaches can substantially distort tail latency observations, and discuss
how the discovery of such problems is inhibited by the common focus
on throughput performance. 
We make a case for purposefully re-designing database benchmarking harnesses
based on these observations to arrive at faithful characterisations of
database performance from multiple important angles.

\keywords{Database benchmarks \and Tail latencies \and Benchmark harness.}
\end{abstract}

\section{Introduction}

Measuring performance is an essential ingredient of evaluating and optimising
database systems, and a large fraction of published
research (\eg,~\cite{10.14778/3364324.3364328,DBLP:phd/dnb/Gessert19,10.14778/2168651.2168654,DBLP:conf/srds/JaimanMQCR18,DBLP:conf/dais/JaimanMR20,DBLP:journals/pvldb/LerschSOL20,DBLP:conf/btw/MauererRFLS21,10.1145/3064176.3064209,garcia2020db2})
is driven
by guidance from the collection of 
benchmarks provided by the Transaction Processing Performance
Council (TPC)~\cite{tpc}, or commercial de-facto standards like the Yahoo! Cloud
Serving Benchmark (YCSB)~\cite{DBLP:conf/cloud/CooperSTRS10}.

These benchmarks usually focus on measuring \emph{throughput} (\ie,
number of operations performed in a given time interval), or latency
(\ie, time from submitting a request to receiving the result, usually
characterised by the 95th or 99th
percentile of the response time distribution).
However, it is known that high latency episodes rarer 
than events in the 99th percentile may
severely impact the whole-system
performance~\cite{10.1145/2408776.2408794}, including important
use-cases like interactive web search~\cite{speed_matters}---even if they
do not receive much attention in standard performance evaluations. 
In this article, we focus on properly characterising tail latencies in
database benchmarking, and unearth shortcomings in popular 
benchmark setups.

We find that tail latencies observed in the ubiquitous TPC-C or YCSB
benchmarks for commonly used databases often fall into the (costly) millisecond~(ms)
range, but are caused by the benchmarking process itself.
Since Barroso~\etal~\cite{Barroso:2017} point out that systemic optimisation efforts
require targeting microsecond~(\(\mu\)s) latencies, aptly termed ``killer microseconds'', it seems evident that non-productive
perturbations that exceed such delays by three orders of magnitude
make it impossible to obtain a faithful characterisation of database
performance.

We show that the popular OLTPBench harness~\cite{DBLP:journals/pvldb/DifallahPCC13,DBLP:conf/cikm/CurinoDPC12},
that is, the software
setting up and executing database benchmarks~\cite{Michael2019},
records latencies that were actually imposed by its own execution 
environment, in particular garbage collection in the Java Virtual Machine
(JVM).
We show that significant noise is caused by the benchmark
harness itself, disturbing the measurements.
However, latencies must be pinpointed to their actual source before targeted improvements can unfold their impact.
If measured latencies are not identified as being caused by the measurement setup, developers will inevitably fail to pin them down, and consequently, to properly address them.

\paragraph{Contributions.}
In this article, we claim the following contributions, 
based on measuring latencies in database query evaluation, using the popular Java-based
benchmark harness OLTPBench~\cite{DBLP:journals/pvldb/DifallahPCC13,DBLP:conf/cikm/CurinoDPC12} with two well-accepted benchmarks (YCSB and TPC-C) on mature database management
systems (MariaDB and PostgreSQL), capturing throughput and tail latencies:

\begin{compactitem}
\item We show that seemingly irrelevant technical details like the
choice of Java Virtual Machine (and even the particular garbage collection mechanism) for the benchmark harness can severely distort tail latencies, increasing maximum latencies by up to several orders of magnitude, while the usually considered quantities median, 95th and 99th percentile of the observed latency distribution remain largely  unperturbed.
\item We carefully separate systemic noise (such as caused by the system software stack) from
the noise intrinsic to the benchmarking harness and the database system. We succeed in
identifying and isolating the latencies introduced by the garbage collector managing
the memory for the benchmark harness.
\item Based on custom-crafted dummy components, we carefully characterise upper \emph{and}
lower bounds for the influence of the measurement infrastructure on the measurement itself,
enabling researchers and developers to distinguish between relevant measurement observations, and
sources of non-productive perturbations caused by the measurement itself.
\item We consistently rely on time-resolved measurements which,
unlike established summary approaches, allow us to discover temporal relations between
events caused by different components of the measurement setup.
\end{compactitem}

\noindent Overall, we systematically build a case for adapting benchmarking harnesses towards faithfully capturing tail latencies.

\paragraph{Structure.}
This paper is organised as follows. We review the preliminaries in Section~\ref{sec:prelims}. We then present our experiments in Section~\ref{sec:experiments}, which we further discuss in Section~\ref{sec:discussion}.
We state possible threats to validity in Section~\ref{sec:threats}, and review related work in Section~\ref{sec:related}. We conclude with Section~\ref{sec:conclusion}.

\section{Preliminaries}
\label{sec:prelims}

\subsection{Database Benchmarks}

TPC-C~\cite{TPCCSpecification}, 
defined in 1992,
is an industry-standard measure for OLTP workloads.
The benchmark 
models order processing in warehouses. 

The Yahoo! Cloud Serving Benchmark (YCSB)~\cite{DBLP:conf/cloud/CooperSTRS10} is an established big data benchmark.
YCSB handles lightweight transactions, as most operations access single records.
This results in low-latency requests, compared to TPC-C.

The No Operation (NoOp) benchmark
provides a simplistic baseline:
To establish lower bounds on achievable latencies, it sends an empty
statement (\eg, the semicolon command for PostgreSQL) that only has to be acknowledged by
the database system, not causing any productive internal processing.
NoOp benchmarks quantify the raw measurement overhead of the benchmark harness, 
and can also be interpreted to represent the minimum client-server \emph{round-trip time} of a
statement.

\subsection{The OLTPBench Benchmark Harness}
A benchmark harness is a toolsuite that provides the functionality to benchmark a software and/or hardware system. 
Typically, a benchmark harness contains components that generate the payload data, execute the database workload, monitor, and collect monitoring data, and even visualise the measured results~\cite{Michael2019}.

For instance, the harness
OLTPBench~\cite{DBLP:journals/pvldb/DifallahPCC13,DBLP:conf/cikm/CurinoDPC12} is a popular~\cite{HRKW13} academic open source project~\cite{OLTPBenchGithub} written in Java.
At the time of writing, the harness implements 19 benchmarks, including the three benchmarks introduced above.
At the time of writing, Google Scholar reports over 280 citations of the full article~\cite{DBLP:journals/pvldb/DifallahPCC13} and the project is rated with over 330 stars on GitHub~\cite{OLTPBenchGithub}, with almost 250 forks.

\subsection{JVM and Garbage Collectors}
\label{sec:JVMGC}
The Java Platform is the specification of a programming language and libraries. The open source OpenJDK~\cite{HotSpotJVM} is a reference implementation since Java~7.

 The \emph{Java Virtual Machine} (JVM) is responsible for all aspects of executing a Java application, including memory management and communication with the operating system.
 The JVM is a specification as well~\cite{JVM16Specification}, with different implementations;
the two most common are the HotSpot JVM~\cite{HotSpotJVM} by OpenJDK and the OpenJ9 JVM~\cite{OpenJ9JVM}, an implementation maintained by the Eclipse Foundation.

The JVM utilises the concept of safepoints.
While implementations differ between JVMs, in general,
an executing thread is in a \emph{safepoint} when its state is well described, that is, all heap objects are consistent. Operations such as executing Java Native Interface (JNI) code require a local safepoint, whereas others, such as garbage collection, require a global safepoint. A global safepoint is a \emph{Stop-The-World (STW)} pause, as all threads have to reach a safepoint and do not proceed until the JVM so decides. The latency of a STW pause is the time once the first thread reaches its safepoint until the last thread reaches its safepoint, plus the time for performing the actual operation that requires an STW pause.

Java is a garbage-collected language.
The \emph{garbage collector} (GC) is a component of the JVM that manages heap memory, 
in particular to remove unused objects~\cite{JVM16Specification}.
These housekeeping tasks can follow different strategies with different optimisation targets, such as optimising for throughput or low latency.
The GC is configured at JVM startup, and additional tuneables can be applied to both, the JVM and the GC of choice.
This allows for optimising a Java application for peak performance based on its environmental conditions, such as hardware aspects or the specific area of use.
However, most GC implementations~\cite{G1GC, ZGC, ShenandoahGC, EpsilonGC, OpenJ9GC} require a global safepoint during their collection phases, which introduces indeterministic latencies.
Azul's C4 GC~\cite{DBLP:conf/iwmm/TeneIW11, DBLP:conf/vee/ClickTW05} overcomes the issue of STW pauses by exploiting read barriers and virtual memory operations, provided by specialised hardware or a Linux kernel module, for continuous and pauseless garbage collection.

\section{Experiments}
\label{sec:experiments}

In the following, we report on the results of our experiments with OLTPBench. 
As a baseline, we execute a minimalist database workload (with the NoOp benchmark), while de-facto disabling the garbage collector (using the HotSpot JVM configured with the Epsilon GC).  This setup is designed to reveal latencies imposed by the benchmark harness
itself on top of payload latencies.
We further configure the harness with special-purpose garbage collectors designed for different scenarios, \eg, which cause only low latencies, and contrast this with the default garbage collectors.

Our experiments are fully reproducible and we refer to our reproduction package\footnote{Zenodo: \url{https://doi.org/10.5281/zenodo.5112729}\\GitHub: \url{https://github.com/sdbs-uni-p/tpctc2021}\label{footnote:repository}} for inspectation and reproduction. The package contains all our measurement scripts, modifications and measured data.

\paragraph{Experimental Setup.}
All experiments are performed with OLTPBench, executing the built-in benchmarks NoOp, YCSB and TPC-C against PostgreSQL and MariaDB.
For Non-Uniform Memory Access (NUMA) awareness, database server and benchmark processes are pinned
to CPUs within the same NUMA node.

\paragraph{Benchmark Configuration.} 
Each benchmark is configured with a ten-second warm up phase, to populate database
caches, buffer pools, etc., followed by a 60-second measurement phase.
The isolation level is set to serialisable, and requests are sent in a closed-loop fashion (a new request will not be sent until the response of the previous request has been received and processed). Requests are issued by ten worker threads in ten parallel connections.

TPC-C is configured with a scale factor of ten, resulting in ten independent warehouses.
Each transaction relates to a specific warehouse using the warehouse ID (primary key).
The warehouse IDs are distributed uniformly over all available worker threads, hence each worker thread executes transactions only on its dedicated warehouse.
This leads to a distribution of one worker per warehouse.
OLTPBench implements TPC-C in ``good faith'' and therefore deviates from the TPC-C specification in some minor details~\cite{DBLP:journals/pvldb/DifallahPCC13}\footnote{For example, TPC-C defines ten terminals (workers) per warehouse and each customer runs through a thinking time at one terminal, which is eliminated by OLTPBench.}.

For YCSB,\colfootnote{OLTPBench is built with the libraries \emph{jaxb-api} and \emph{jaxb-impl} in version 2.3.0, which leads to a NullPointerException with Java versions $\geq 9$. This issue is resolved in the libraries with version 2.3.1, to which we updated.} we use a scale factor of 1,200, resulting in a table with 1.2 million records.
The workload distribution is as follows: 50\% read, 5\% insert, 15\% scan, 10\% update, 10\% delete and 10\% read-modify-write transactions, while all but scan access a single record based on the primary key.
The primary key selection is based on a Zipfian distribution. All these settings are defaults in OLTPBench.

In the current implementation of OLTPBench, the NoOp benchmark is only supported by PostgreSQL.
In case of an empty query, PostgreSQL will acknowledge the query and report successful execution. However,
MariaDB reports an empty query error, which results in a Java runtime exception on the side of the benchmark, which, in turn, results in different code paths, compared to PostgreSQL.
To promote comparable behaviour, we enhanced both, MariaDB and OLTPBench:
In OLTPBench, we disabled explicit \emph{commit}s after transactions.\colfootnote{A commit after an empty query does not have any effects on execution.}
Additionally, we enhanced MariaDB to interpret the empty statement ``\texttt{;}'' internally as comment~(\texttt{--}), that is, as a NoOp. The modifications are part of our reproduction package (see Footnote \ref{footnote:repository}).

\paragraph{Java and GC Settings.}
To run OLTPBench, we use Java version 16.0.1 (OpenJDK).
We measure with both the HotSpot JVM and OpenJ9 JVM.
For the HotSpot JVM, we use the garbage collectors G1 (default)~\cite{G1GC}, Shenandoah~\cite{ShenandoahGC}, ZGC~\cite{ZGC}, and Epsilon~\cite{EpsilonGC}. The latter is a pseudo garbage collector, it leaves all objects in memory and performs no garbage collection at all.
For OpenJ9, we used gencon (default)~\cite{OpenJ9GC} and metronome~\cite{OpenJ9GC}.
Table~\ref{tab:gcs} provides an overview of the strategies of the GCs used.
We chose the default GC for HotSpot JVM and OpenJ9 JVM as they are probably the starting point for most measurements done with a Java application. In addition, we choose all low latency GC strategies with short STW pauses for precise latency measurements. 

\begin{wraptable}[18]{R}{\linewidth/2}
\vspace*{-3em}\caption{Garbage collectors for the HotSpot JVM and OpenJ9 JVM, and design goals.}\label{tab:gcs}

\begin{tabularx}{\linewidth}{llX}
\toprule
JVM & GC & Design\\\midrule
 \parbox[t]{2mm}{\multirow{8}{*}{\rotatebox[origin=c]{90}{HotSpot}}} & G1 & Balance  throughput and latency.\\ 
                         & Z          & Low latency.\\ 
                         & Shenandoah & Low latency.\\ 
                         & Epsilon  & Experimental setting: No GC tasks are performed, except for increasing heap.\\\midrule
 \parbox[t]{2mm}{\multirow{4}{*}{\rotatebox[origin=c]{90}{OpenJ9}}}  & gencon  & Transactional applications with short-lived objects.\\
                         & metronome & Low latency.\\
\bottomrule
\end{tabularx}
\end{wraptable}

\paragraph{Experiment Execution.}
By default, OLTPBench sets a maximum heap size of $\SI{8}{\gibi\byte}$ (JVM option \texttt{-Xmx8G}), which we also used for our experiments, with the exception of Epsilon GC.
As the Epsilon GC does not perform garbage collection, we enlarge the heap size accordingly:
In total, the Epsilon GC requires $\SI{180}{\gibi\byte}$ of heap space of which $\SI{160}{\gibi\byte}$ were pre-allocated upon startup. 
During the 60-second measurement, $\SI{160}{\gibi\byte}$ heap space were sufficient, so no latencies were introduced due to an increase of the heap.
The remaining $\SI{20}{\gibi\byte}$ of heap space that were not pre-allocated, but reserved,
were required by OLTPBench for the YCSB benchmark to create result files.

To capture latencies introduced by the JVM, 
and to confirm our hypothesis, we consult a second dataset.
We exploit the unified logging mechanism, introduced in HotSpot JVM for Java~9~\cite{UnifiedJVMLoggingHotSpot}.
It allows for logging all safepoint events, including those for garbage collection.
OpenJ9 JVM also provides unified logging, but only records garbage collection events and thus no safepoint events~\cite{UnifiedJVMLoggingOpenJ9}.
We mine the log for safepoint events, or in case of OpenJ9 JVM for GC events, and interpret these latencies as overhead caued by the JVM.

\paragraph{Execution Platform.}
All measurements are performed on a Dell PowerEdge R640, with two Intel Gold 6248R CPUs (24 cores per CPU, \SI{3.0}{\giga\hertz}) and $\SI{384}{\giga\byte}$ of main memory.
To avoid distortions from CPU frequency transitions, we disable Intel\textsuperscript{\textregistered\texttrademark} Turbo Boost\textsuperscript{\textregistered\texttrademark}, and operate all cores in the performance P-State for a permanent core frequency of \SI{3.0}{\giga\hertz}.
We disable simultaneous multithreading (SMT) since it causes undesired side-effects on low-latency real-time systems~\cite{Mauerer:2010} due to resource contention.

To avoid cross-NUMA effects, OLTPBench as well as database server processes execute on 22 cores of the same NUMA node: OLTPbench and the database server (either MariaDB or PostgreSQL) each execute exclusively on eleven cores.
This ensures one worker of the benchmark, which is pinned to a dedicated CPU, is connected to one worker of the database, which is pinned to a dedicated CPU as well.
The remaining two cores of the NUMA node are reserved for the remaining processes of the system.

The server runs Arch Linux with Kernel version 5.12.5.
The benchmark as well as the database server are compiled from source.
For OLTPBench, we use the version of git hash \#6e8c04f, for PostgreSQL Version 13.3, git hash \#272d82ec6f and for MariaDB Version 10.6, git hash \#609e8e38bb0.

\subsection{Results}
\label{sec:results}

Our evaluation with MariaDB and PostgreSQL shows that the results are virtually independent of the DBMS. Hence, we focus on presenting the results for MariaDB.
For PostgreSQL, we refer to our reproduction package (see Footnote~\ref{footnote:repository}).

We follow a top-down approach:
We first measure latencies on the level of single transactions, under variation of different JVM and GC configurations with the NoOp, YCSB, and TPC-C benchmarks. We then characterise the latency distributions, and systematically investigate the latency long tail.

\subsubsection{Throughput.}

\begin{figure}[tb]
    \includegraphics[width=\textwidth]{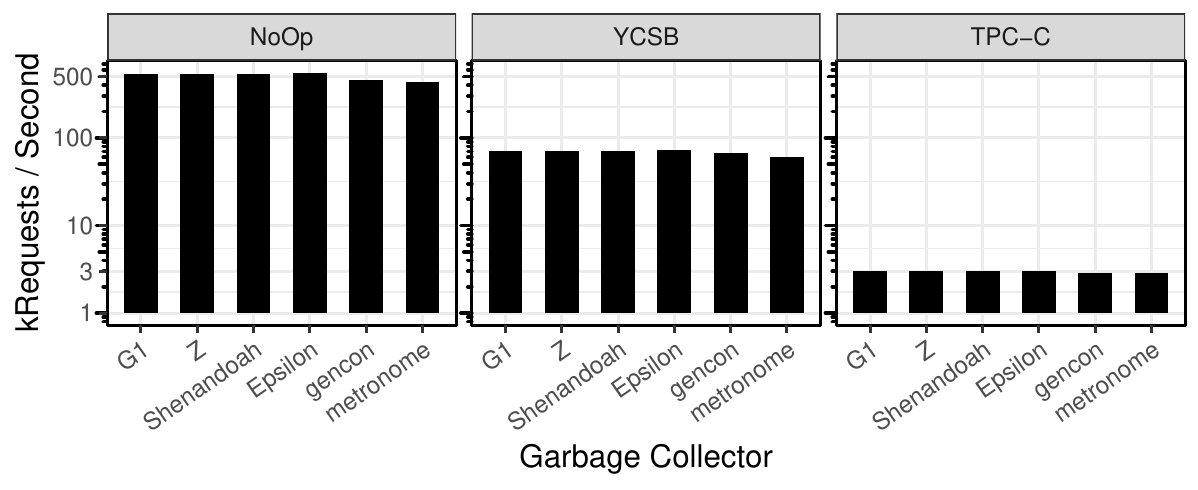}
    \vspace*{-2em}
    \caption{Database throughput  for MariaDB, in thousand requests per second for benchmarks NoOp, YCSB and TPC-C, and different JVM/GC configurations of  OLTPBench. Throughput is affected marginally by the choice of JVM, but not the GC.}
    \label{fig:rps}
\end{figure}

 Figure~\ref{fig:rps} shows the throughput measured in thousand requests per second, for different benchmarks, JVMs, and GCs. For TPC-C, throughput is commonly reported as NewOrder transactions per minute (tpmC), but we deviate for a better comparability between the different benchmarks. 
 
For all three benchmarks, Figure~\ref{fig:rps} reports 
a difference in performance between the two JVMs: Compared to OpenJ9 JVM, HotSpot JVM has about 17\% -- 28\% more throughput for the NoOp benchmark,  about 5\% -- 18\% for YCSB, and only about 5\% more for the TPC-C benchmark (note the log-scaled vertical axis). 

Naturally, the choice of JVM for OLTPBench has a stronger influence for benchmarks in which comparatively little time is spent on the database side. To put this in context: By executing the NoOp benchmark with about 500k requests per second, we spend much more time in the process of OLTPBench compared to the TPC-C benchmark with only 3k requests per second.

\subsubsection{Latency distribution.}

\begin{figure}[htb]
    \includegraphics[width=\textwidth]{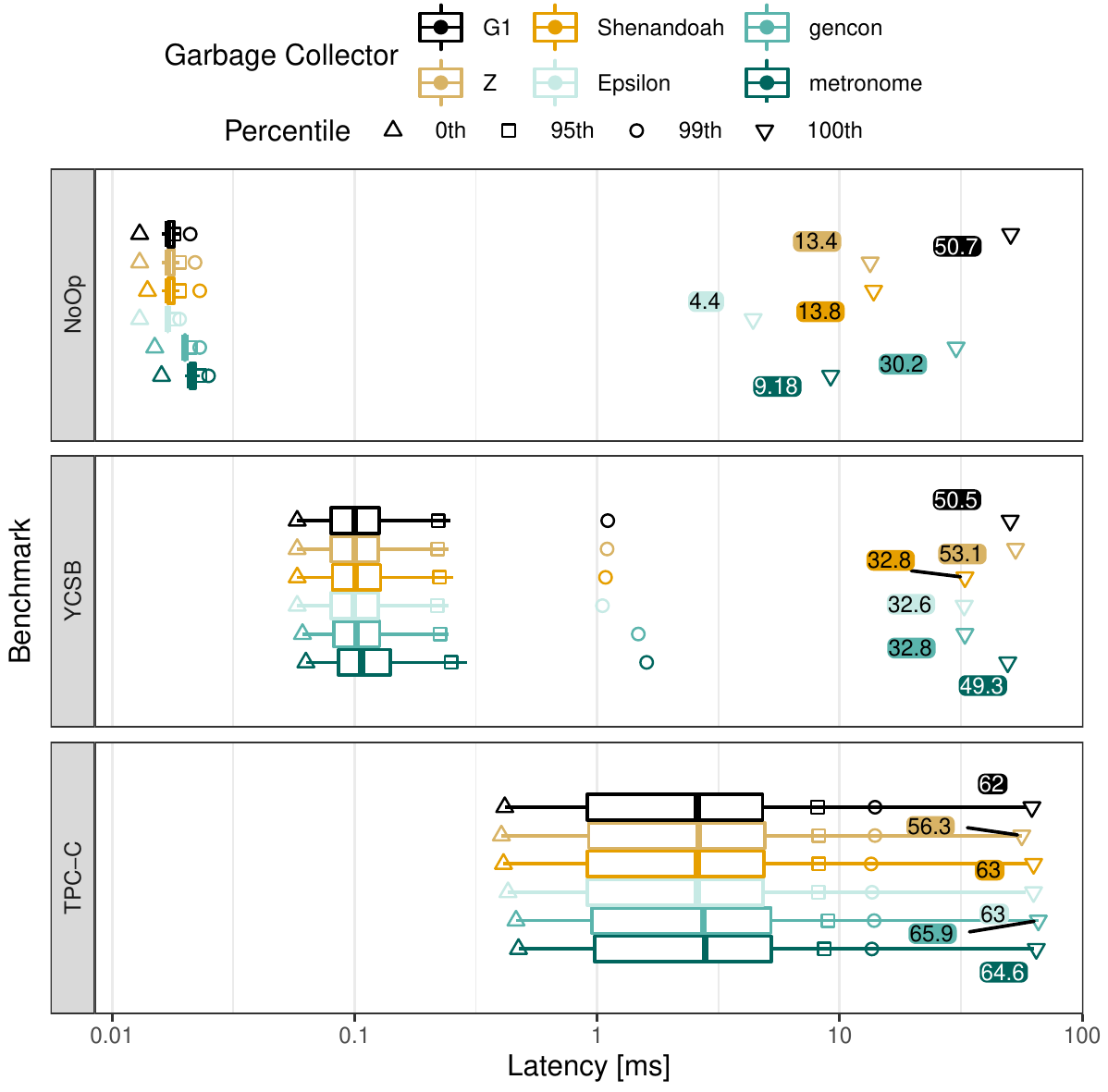}
    \vspace*{-2em}
    \caption{The latency distributions measured by OLTPBench for three benchmarks, visualised as box plots. 
    Key percentiles 
    are highlighted.}
    \label{fig:percentiles}
\end{figure}

The distribution of latencies reported by OLTPBench is visualised by box-plots in
Figure~\ref{fig:percentiles}. The minimum (0th percentile) and maximum (100th percentile)
latencies, as well as the 95th and 99th latency percentiles, are marked separately. The
absolute value of the maximum latency is also labelled.

Overall, there is little variation in the median latencies within a benchmark.
Comparing the median latencies of the two JVMs for the NoOp benchmark, we see a lower median latency for HotSpot JVM than for OpenJ9 JVM: All GCs of HotSpot JVM have a median latency of \SI{0.017}{\milli\second}, while gencon has \SI{0.020}{\milli\second} and metronome \SI{0.021}{\milli\second}, based on about 500k requests per second. As the NoOp benchmark generates only a small load on the database side, the maximum latencies reported are candidates for latencies introduced by the Java environment of the benchmark harness. As expected, Epsilon GC has the lowest maximum latency for the NoOp benchmark.

The YCSB benchmark shows strong variance in maximum latencies, depending on the garbage collector used. For GCs G1, Z and metronome, a maximum latency of around \SI{50}{\milli\second} is recorded, whereas the other GCs display a maximum latency of about \SI{30}{\milli\second}.
We inspect these latencies more closely in the following. Nevertheless, the distribution of the latencies is close to uniform for all six garbage collectors, except the 99th percentile latency: Observations for OpenJ9 exceed those for HotSpot. 

The different JVM/GC configurations  result in near-identical latency distributions for the TPC-C benchmark: Due to the larger share of time spent on the database side (compared to the other benchmarks), the latencies introduced by the benchmark harness do not weigh in as much in comparison.

\subsubsection{Latency time series.}
\begin{figure}[htb]
    \includegraphics[width=\textwidth]{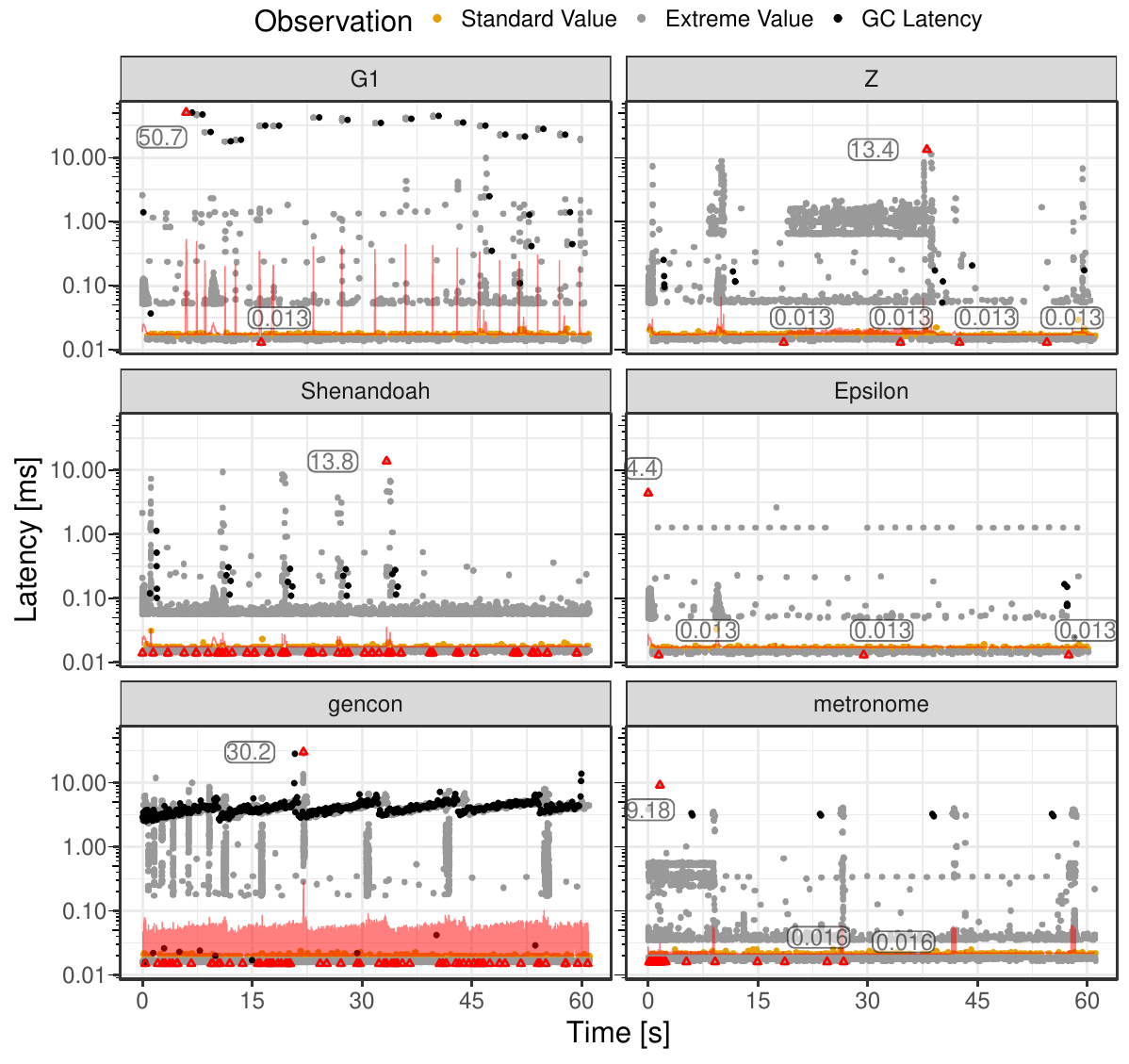}
    \vspace*{-2.5em}\caption{Latency time series of the NoOp benchmark. Minimum and maximum latencies measured with OLTPBench are marked by red, labelled triangles. Grey dots represent extreme values, 
    ochre dots (down-sampled) standard observations.
    Latencies from the JVM log file are superimposed in black. The red line shows the sliding mean window.
    }
    \label{fig:latenciesNoop}
\end{figure}

\begin{figure}[htb]
    \includegraphics[width=\textwidth]{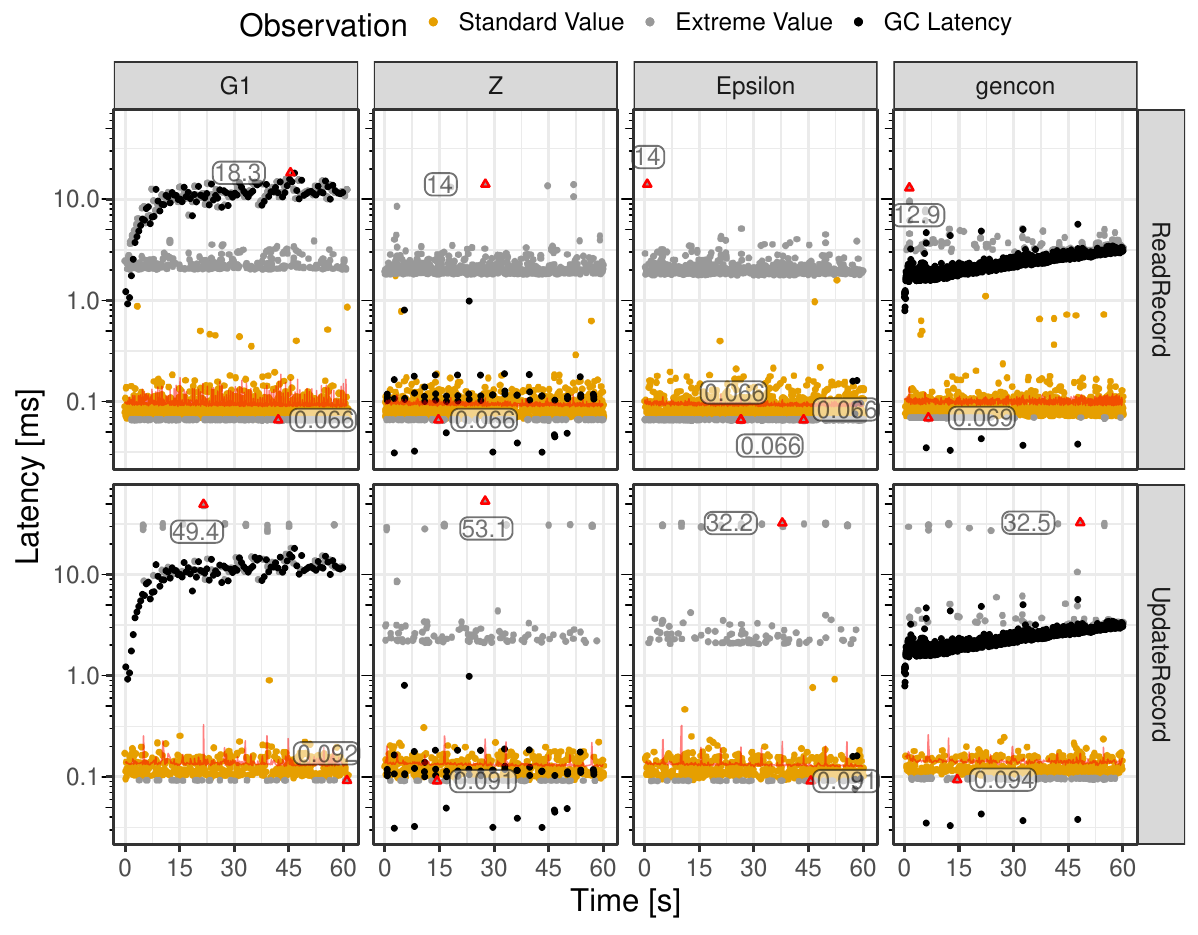}
    \vspace*{-2.5em}
    \caption{Latency time series of the YCSB benchmark for read (ReadRecord) and write (UpdateRecord) transactions. 
    Labels and colours as in Figure~\ref{fig:latenciesNoop}.
    }
    \label{fig:latenciesYcsbQueries}
\end{figure}

\begin{figure}[htb]
    \includegraphics[width=\textwidth]{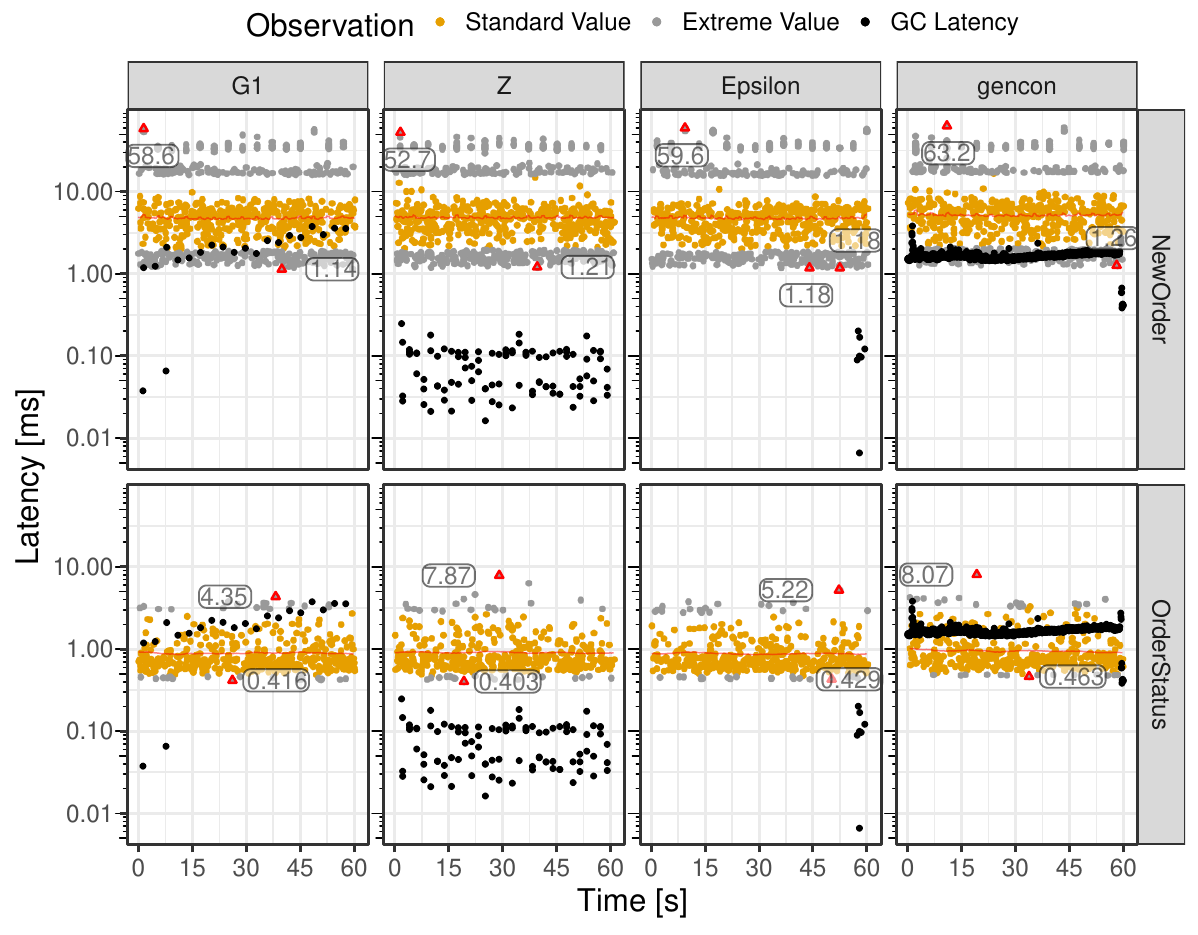}
    \vspace*{-2.5em}
    \caption{Latency time series of the TPC-C benchmark for read (OrderStatus) and write (NewOrder) transactions.
       Labels and colours as in Figure~\ref{fig:latenciesNoop}.
    }
    \label{fig:latenciesTpccQueries}
\end{figure}

Figures~\ref{fig:latenciesNoop} through~\ref{fig:latenciesTpccQueries}
show time series plots for the benchmarks NoOp, YCSB, and TPC-C.
Red, labelled triangles mark minimum and maximum latencies, as observed by  OLTPBench. In order to prevent overplotting, we downsampled the latencies except for extreme values.
Ochre dots represent sampled, normal observations. 
A latency is considered an extreme value and displayed as grey dot if it is outside a pre-defined range. We define benchmark-specific sampling rates and extreme value ranges, as detailed below. 
Latency fluctuations smoothed by a sliding window covering 1,000 data points are shown in red.

Superimposed black dots represent latencies extracted from the JVM logs. Since
randomised, mixed workloads do not allow us to associate given latencies with specific queries,
we visualise all JVM latencies.

The time series plots for the YCSB and TPC-C benchmark are provided for selected queries only. We refer to our reproduction package (see Footnote~\ref{footnote:repository}) for the full set of charts, which confirm our observations.
Similarly, we do not visualise the Shenandoah GC as it behaves similar to the Z GC and metronome GC, with a similar latency pattern as gencon GC. 

\paragraph{NoOp Benchmark.}
The latency time series for the NoOp benchmark is shown in Figure~\ref{fig:latenciesNoop}.
To arrive at meaningful visualisations, we apply a sampling strategy that avoids overplotting
for ``standard'' observables by only using 0.001\% of the recorded values in between the
0.025th and 99.975th percentile. However, we show the full collection of observations outside this
interval.

OLTPBench, executed with the Epsilon GC, shows that this setup has the lowest latency
possible, and the JVM is only active for a short time at the very end. This measurement shows
that regular outliers in the range of about \SI{1}{\milli\second} occur regularly in
the time series, and can be used as a reference for comparison with other GCs. The measurement of the GC G1 shows that the GC causes the maximum latency and the tail of the latencies.
Almost each latency higher than \SI{10}{\milli\second} was introduced by the GC because the latencies reported by OLTPBench and these reported by the JVM display the same pattern and match each other.

\paragraph{YCSB Benchmark.}
We show the time series latency of the YCSB benchmark in Figure~\ref{fig:latenciesYcsbQueries}. We report the latencies measured by the G1, Z, Epsilon and gencon GC and selected the two transaction types ReadRecord (read transaction) and UpdateRecord (write transaction). We used a sampling rate of 0.05\% (ReadRecord) and 0.1\% (UpdateRecord) for standard values and the 99.975th and 0.025th latency percentile are the limits for a latency to be marked as an extreme value.

The Epsilon GC is again the reference and except for the maximum latency, all tail latencies fall into the interval between \SI{1}{\milli\second} and \SI{5}{\milli\second}. By comparing the ReadRecord latencies reported from OLTPBench and from the JVM, again the G1 GC is responsible for the tail latencies occurring in this transaction. The write transaction shows a similar behaviour, but here outliers on the database side are responsible for the maximum latency, nevertheless again the G1 GC latency defines the tail.

\paragraph{TPC-C Benchmark.}
The time series latency of TPC-C is shown in Figure~\ref{fig:latenciesTpccQueries}. The sampling
rate of standard values of the NewOrder transaction is set to 0.5\% and for OrderStatus to 5\%.
Extreme values are marked as such if they exceed the 99.75th percentile or subceed the 0.25th
percentile. 

For this particular benchmark, the influence of the JVM is negligible. The
transactions, especially write transactions, are so heavyweight that the processing time inside the database
substantially exceeds the benchmark overhead. Furthermore, due to the low number of requests per second (about 3k),
only a limited amount of intermediate, temporary objects that require garbage collection are created in the first
place. The same applies to read transactions. 

\section{Discussion}
\label{sec:discussion}

Our experiments show that for the popular benchmark harness OLTPBench, the choice of the execution environment (JVM and its GC strategy) substantially impacts  (tail) latency measurements. 
By super-imposing the latencies extracted from JVM log files on the latency time series reported by OLTPBench, we make this connection visually apparent, and 
show that different GC strategies translate to different temporal patterns. 
By setting up a baseline experiment, with garbage collection de-facto disabled, and a minimalist database workload, we can successfully establish a lower bound on
non-productive latency contributions.

Naturally, for lightweight database workloads (in our experiments, YCSB), this non-productive overhead is more noticeable in relation to the actual query execution time.

Interestingly, while researchers and practitioners optimise latencies in the realm of microseconds~\cite{Barroso:2017}, the latencies imposed by the benchmark harness reach the ballpark of milliseconds. 
Evidently, this factor of one thousand proves these effects are non-negligible, and deserve careful consideration in any evaluation, albeit most published research
neglects this issue so far.

Our observations are replicable for PostgreSQL. We provide the data and full set of plots, along with a reproduction package (see Footnote~\ref{footnote:repository}).

\section{Threats to Validity}
\label{sec:threats}

We applied great care and diligence in analysing the garbage collector logs.
Yet as the logging mechanisms differ between JVMs, we must deal with some level of uncertainty: 
The HotSpot JVM logs all safepoints (including, but not restricted to garbage collection events), whereas the OpenJ9 JVM logs only the garbage collection events. 
As the GC events dominate the logged safepoint events, we treat the reported latencies in both logs uniformly.
In addition, we do not distinguish between local and global safepoints, as local safepoints can also block a thread.

Further, the latencies reported by OLTPBench and the latencies logged by the JVM are two distinct sources of data. As usual, data integration brings uncertainties, so the latencies displayed from OLTPBench and JVM might be minimally shifted in time in the time series plots.
In summary, we consider the above threats to the validity of our results as minor.

One further threat to validity is that we only focus on the Java environment of the benchmark harness, but no other possible sources of systemic noise (such as caused by the hardware). We have diligently configured the execution platform to eradicate sources of noise
(\eg, by disabling SMT).
Moreover systemic noise is typically in the range of micro seconds~\cite{Barroso:2017}. Since the harness-induced latencies are in the millisecond range (exceeding them by a factor of one thousand, and clearly traceable back to the harness), we may dismiss this threat.

\section{Related Work}

\label{sec:related}

Databases, as core components of data-intensive systems, are important
contributors to system-wide tail latency effects. Likewise, they 
have started to enjoy increasing 
popularity in real-time scenarios like sensor networks
or IoT~\cite{paparrizos2021vergedb,garcia2020db2}, where the most crucial service indicator is determinism and \emph{maximum} latency, instead of
average-case performance. Care in controlling excessive
latencies must be exercised in all cases.

As modifiable open source components are increasingly used in building systems~\cite{HRKW13,Ramsauer:2016},
several measures addressing software layers above, inside and below the DBMS
have been suggested to this end.
For instance, optimising end-to-end
communications~\cite{DBLP:phd/dnb/Gessert19,DBLP:conf/btw/ScherzingerKS09,patent_e2e}
tries to alleviate the issue from above. 
%
Specially crafted real-time databases~\cite{DBLP:conf/rtdb/2001},
novel scheduling algorithms in scheduling/routing queries~\cite{DBLP:conf/srds/JaimanMQCR18,DBLP:conf/dais/JaimanMR20,10.1145/3064176.3064209}, transactional concepts~\cite{10.1145/3447786.3456238},
or query evaluation strategies~\cite{10.14778/2168651.2168654,DBLP:journals/pvldb/UnterbrunnerGAFK09}, work from inside the database.
%
Careful tailoring of the whole software stack from OS kernel to DB engine\cite{DBLP:conf/btw/MauererRFLS21,10.1145/2670979.2670988},
or crafting dedicated operating systems~\cite{DBOS,mueller:2019:sfma,Muehlig2020,DBLP:journals/debu/Giceva19,DBLP:conf/cidr/GicevaSSAR13,Giceva:2016}
to leverage the advantages of modern hardware in database system engineering
(\eg,~\cite{DBLP:journals/pvldb/LerschSOL20,DBLP:conf/icde/FentRKL0K20}),
contribute to solutions from below the database.

In our experiments, we execute the well-established YCSB and TPC-C benchmarks, which are supported by the OLTPBench harness out-of-the-box.
However, further special-purpose benchmarks have been proposed specifically for measuring tail latencies, as they arise in embedded databases on mobile devices~\cite{DBLP:conf/tpctc/NuessleKZ19}. This contribution is related to our work insofar as the authors 
also unveil sources of measurement error, however, errors  that arise when measuring the performance of embedded databases at low throughput.

There are further benchmark harnesses from the systems research community that specifically target tail latencies.
These  capture latencies across the entire systems stack, while in the database research community, we benchmark the database layer in isolation.
Harnesses such as TailBench~\cite{7581261} or TreadMill~\cite{7551414} define such complex application scenarios (some involving a database layer).

In its full generality, the challenge of benchmarking Java applications, including jitter introduced by garbage collection, is discussed in~\cite{10.1145/1297105.1297033}.
We follow the best practices recommended in this article, as we classify outliers as either systematic or probabilistic. We present the latency distributions clearly, and have carried out sufficiently many test runs.
Different from the authors of~\cite{10.1145/1297105.1297033}, we do not apply hypothesis tests, since we are mostly interested in latencies maxima, rather than confidence in averaged values.

It has been reported that database-internal
 garbage collection~\cite{10.14778/3364324.3364328,DBLP:journals/pvldb/LerschSOL20}
 can also cause latency spikes, which might however be seen as part of productive operation. Our work considers the effects of garbage collection inside the
 test harness, rather than the database engine.

\section{Conclusion and Outlook}
\label{sec:conclusion}

Tail latencies in database query processing can manifest as acute pain points. 
To address their possible causes, we must be able to faithfully measure them.
Our work shows that Java-based benchmark harnesses serve well for measuring database throughput,
or 95th or 99th percentile latencies. However, these benchmarks
can significantly impact the capturing of extreme tail latencies:
The choice of JVM and garbage collector in the harness is a non-negligible source of indeterministic noise.
For database workloads composed of low-latency queries (\eg, as in the YCSB benchmark), we risk distorted
measurements which can lead us to chase ghosts in database systems engineering, and prevent an accurate and faithful
characterisation of important extreme events. We find that future efforts in evaluating database performance for real-time and
large-scale computing scenarios should put more effort into understanding and controlling such effects.

%
%
\bibliographystyle{splncs04}
\bibliography{references}
\end{document}